\documentclass[preprint,nofootinbib,superscriptaddress,amssymb]{revtex4}
\usepackage{amsmath} 
\usepackage{verbatim} 
\usepackage{graphicx} 
\usepackage{bbold} 
\usepackage{multirow} 

\makeatletter
\renewcommand*\env@cases[1][1.2]{%
  \let\@ifnextchar\new@ifnextchar
  \left\lbrace
  \def\arraystretch{#1}%
  \array{@{}l@{\quad}l@{}}%
} 
\makeatother

\begin{document}

\title{Do Cosmological Perturbations Have Zero Mean?}

\author{Cristian Armendariz-Picon}
\affiliation{Department of Physics, Syracuse University, Syracuse, NY 13244-1130, USA}

\begin{abstract}
A central assumption in our analysis of cosmic structure is that cosmological perturbations have zero ensemble mean. This property is one of the consequences of statistically  homogeneity, the invariance of  correlation functions under spatial translations. In this article we explore whether cosmological perturbations indeed have zero mean, and thus test one aspect of statistical homogeneity. We carry out a classical test of the zero mean hypothesis  against a class of alternatives in which perturbations have non-vanishing means, but homogeneous and isotropic covariances. Apart from Gaussianity, our test does not make any additional assumptions about the nature of the perturbations and is thus rather generic and model-independent. The test statistic we employ is essentially  Student's $t$ statistic, applied to appropriately masked, foreground-cleaned cosmic microwave background anisotropy maps produced by the WMAP mission. We find evidence for a non-zero mean in a particular range of multipoles, but the evidence against the zero mean hypothesis goes away when we correct for multiple testing.   We also place constraints on the mean of the temperature multipoles as a function of angular scale.  On angular scales smaller than four degrees, a non-zero mean has to be at least an order of magnitude smaller than the standard deviation of the temperature anisotropies. 

\end{abstract}

\maketitle

\section{Introduction}

The cosmological principle is one of the cornerstones of modern cosmology. Roughly speaking, the principle  states that the universe is homogeneous and isotropic on large scales. Although large-scale homogeneity and isotropy were initially postulated, in recent decades the  principle has received mounting experimental support, and today there is little doubt about its validity. The cosmological principle has a slightly preciser formulation, which states that a perturbed Friedman-Robertson-Walker metric provides an accurate description of the universe. Thus, according to the cosmological principle, the universe is well-described by the perturbed spacetime metric
\begin{equation}
 	ds^2=a^2(\eta)\left[-(1+2\Phi(t,x))d\eta^2+(1-2\Psi(t,x))d\vec{x}^2\right],
\end{equation}
with sufficiently small (scalar) perturbations at long wavelengths
\begin{equation}
	\Phi(t, \vec{k})\ll 1,  \quad \Psi(t, \vec{k})\ll 1.
\end{equation}
But apart from that, the principle has nothing to say about the properties of these perturbations.

Many of the advances in modern cosmology consist in the characterization  of the  metric perturbations $\Phi(\vec{x})$ and $\Psi(\vec{x})$. Though not often explicitly emphasized, one of the key assumptions is that these perturbations are just a particular realization of a random process in a statistical ensemble. Hence, we do not really try to describe the actual perturbations $\Phi(t,\vec{x})$ and $\Psi(t,\vec{x})$; our goal is to characterize the statistical properties of the random fields $\hat{\Phi}(t,\vec{x})$ and $\hat{\Psi}(t,\vec{x})$.  

Let $\hat{\phi}(t,x)$ denote any random field in the universe, such as the metric perturbations considered above, or the energy density of any of the components of our universe. The statistical properties of the random field are uniquely specified by its probability distribution functional. It turns to be simpler however to study the moments of the field ${\langle \hat{\phi}(\vec{x}_1) \cdots \hat{\phi}(\vec{x}_n)\rangle},$  where $\langle \cdots \rangle$ denotes ensemble average, and all the fields are evaluated at a common but arbitrary time $t$, which we suppress for simplicity. 

The cosmological principle has  formal counterparts in the properties of the perturbations, though, as we emphasized above, the cosmological principle itself only requires the actual perturbations in our universe to be small. We say that a random field $\hat{\phi}$ is statistically homogeneous (or stationary), if all its moments are invariant under translations,
\begin{equation}
	\left\langle \hat{\phi}(\vec{x}_1)\cdots \hat{\phi}(x_n) \right\rangle
	=\langle  \hat{\phi}(\vec{x}_1+\vec{T})\cdots  \hat{\phi}(\vec{x}_n+\vec{T})\rangle,
	 \quad \forall \vec{T}\in\mathbb{R}^3, \, \forall n\in \mathbb{N}.
\end{equation}
In some cases, statistical homogeneity may apply only to some field moments. The random field is \emph{stationary in the mean} if 
\begin{equation}\label{eq:stationary mean}
 \langle \hat{\phi}(\vec{x}) \rangle	= \langle \hat{\phi}(\vec{x}+\vec{T}) \rangle \quad 
 \forall \vec{T}\in\mathbb{R}^3,
\end{equation}
and it  is \emph{stationary in the variance} if 
\begin{equation}
	\left\langle \Delta\hat{\phi}(\vec{x}_1)\,  \Delta\hat{\phi}(\vec{x}_2)\right\rangle=
	\left\langle \Delta\hat{\phi}(\vec{x}_1+\vec{T})\, 
	\Delta\hat{\phi}(\vec{x}_2+\vec{T})\right\rangle \quad  \forall \vec{T}\in\mathbb{R}^3,
\end{equation}
where we have defined $\Delta\hat \phi\equiv \hat\phi-\langle \hat\phi\rangle$.  If the random field is Gaussian, the one- and two-point functions uniquely determine all the remaining moments of the field. A Gaussian random field stationary in the mean and in the variance is hence automatically fully stationary.

Parallel definitions apply to the properties of the perturbations under rotations. In particular, we say that a  random field $\hat{\phi}$ is  \emph{isotropic  in the mean} if
\begin{equation}\label{eq:isotropic mean}
	\langle \hat{\phi}(\vec{x})\rangle =
	 \langle \hat{\phi}\left(\vec{o}+R\cdot(\vec{x}-\vec{o}\,)\right) \rangle \quad \forall\vec{o}\in \mathbb{R}^3,\, \forall R\in SO(3).
\end{equation}
Analogously, a random field is \emph{isotropic in the variance} if
\begin{equation}
	\left\langle 
	\Delta\hat{\phi}(\vec{x}_1) \, \Delta\hat{\phi}(\vec{x}_2)
	\right\rangle= 
	\left\langle 
	\Delta\hat{\phi}\left(\vec{o}+R\cdot(\vec{x_1}-\vec{o}\,)\right) \,
	\Delta\hat{\phi}\left(\vec{o}+R\cdot(\vec{x_2}-\vec{o}\,)\right)\right\rangle
	\quad \forall\vec{o}\in \mathbb{R}^3,\, \forall R\in SO(3).
\end{equation}
Since  there is always a rotation that maps $\vec{x}$ to $\vec{x}+\vec{T}$, and because any two points related by a rotation always differ by a translation, equations (\ref{eq:stationary mean}) and (\ref{eq:isotropic mean}) imply that homogeneity and isotropy in the mean are equivalent.  But homogeneity in the variance  \emph{does not} imply isotropy in the variance, though the converse is true \cite{ArmendarizPicon:2005jh},
\begin{equation}\label{eq:equivalenceA}
	\text{Isotropy in the variance}\Rightarrow \text{Homogeneity in the variance}.
\end{equation}

Homogeneity and isotropy in the mean have an important consequence: Equations (\ref{eq:stationary mean}) or (\ref{eq:isotropic mean}) immediately imply that the expectation of a stationary field is constant,
\begin{equation}
	\langle \hat{\phi}(\vec{x})\rangle = \text{const},
\end{equation}
and, conversely, any random field with constant mean is homogeneous and isotropic in the mean.  Because, by definition, cosmological perturbations always represent  deviations from a homogeneous and isotropic background, it is then always possible to assume that the constant value of their mean is zero, if they happen to be stationary. For example, in perturbation theory we write the total energy density $\rho$ as a background value $\rho_0$ plus  a perturbation $\delta\rho$,
\begin{equation}\label{eq:split}
	\rho=\rho_0(t)+\delta\rho(t,x).
\end{equation}
This split into a background value and a perturbation is essentially ambiguous, unless we specify what the background actually is. In cosmology, what sets the background apart from the perturbations is symmetry. Because of the cosmological principle, the background energy density $\rho_0$ is \emph{defined} to be homogeneous. Hence, if the constant mean of the stationary random field $\delta\rho$ is not zero, we may redefine our background and perturbations by
\begin{equation}\label{eq:redefinition}
	\rho_0 \to \tilde{\rho}_0\equiv  \rho_0+\langle \delta\rho\rangle,
	\quad
	\delta\rho\to \delta\tilde{\rho}\equiv \delta\rho-\langle \delta\rho \rangle,
\end{equation}
without affecting the overall value of the energy density, $\rho\to \tilde{\rho}=\rho$.  In this case the redefined perturbation $\delta\tilde{\rho}$ has zero mean, while the redefined background  $\tilde{\rho}_0$ is still space-independent.

It is important to recognize that cosmological perturbations can be assumed to have zero mean if and only if their mean is a constant.  Consider again the example of the energy density (\ref{eq:split}), but now assume that $\delta\rho$ is not stationary.  Although the redefinitions (\ref{eq:redefinition}) allow us to set the mean of the perturbations $\delta\tilde\rho$ to zero, the redefined background $\tilde{\rho}_0$ is inhomogeneous in this case, in contradiction with our definition of the background density $\rho_0$ in equation (\ref{eq:split}). Therefore, we conclude that homogeneity in the mean, isotropy in the mean and zero mean are all equivalent,
\begin{equation}\label{eq:equivalenceB}
\text{Zero mean}\Leftrightarrow \text{Homogeneity in the mean}\Leftrightarrow \text{Isotropy in the mean}.
\end{equation}

Homogeneity and isotropy in the variance also have important implications \cite{ArmendarizPicon:2005jh}. If a random field is stationary in the variance, its two point function in momentum space has to be proportional to a delta function,
\begin{equation}
	\langle \Delta\phi(\vec{k}_1) \Delta\phi (\vec{k}_2)\rangle\equiv (2\pi)^3  \delta(\vec{k}_1+\vec{k}_2)
	\frac{2\pi^2 \mathcal{P}_\phi(\vec{k}_1) }{k^3},
\end{equation}
and if the variance is isotropic, the power spectrum $\mathcal{P}_\phi$ can only depend on the magnitude of $\vec{k}$,
\begin{equation}
	\mathcal{P}_\phi(\vec{k})=\mathcal{P}_\phi(k).
\end{equation}

Based on the  equivalences (\ref{eq:equivalenceA}) and (\ref{eq:equivalenceB}), there are hence six possible different combinations of the statistical  properties of the primordial perturbations, which we list in table \ref{tab:cases}. Hypothesis $H_0$ describes the standard assumption that underlies most  analyses of cosmological perturbations, and  case $H_1$ describes what is usually known as a violation of statistical isotropy. In this article we focus on violations of the  zero mean hypothesis, cases $H_3$ through $H_5$.   Our goal is to test the standard assumption $H_0$ against one of its non-zero mean alternatives.

\begin{table}
\begin{tabular}{|c|c|c|c|}
\hline
\,  \multirow{2}{*}{Hypothesis} \, &{}\,  Mean  \, {} & \multicolumn{2}{c|}{Variance} \\
  \cline{2-4}
  & Zero & {}\, Homogeneous\, {} &  {}\,  Isotropic    {}\, \\
 \hline 
$H_0$ & yes & yes & yes \\
$H_1$ & yes & yes & no \\
$H_2$ & yes & no & no \\
\hline
$H_3$ & no & yes & yes \\
$H_4$ & no & yes & no \\
$H_5$ & no & no & no\\
\hline
\end{tabular} 
\caption{The six possible different combinations of statistical properties of the primordial perturbations. We are concerned here with the mean and variance alone. \label{tab:cases}}
\end{table}

\section{Temperature Anisotropies}

At present, the arguably cleanest and widest window on  the primordial perturbations in our universe stems from the temperature anisotropies observed in the Cosmic Microwave Background Radiation (CMB). Hence, if we want to test whether cosmological perturbations have zero mean, we need to explore how these temperature anisotropies are related to the random fields that we have considered in the introduction.  

\subsection{Harmonic Coefficients}

In a homogeneous and isotropic universe, different Fourier modes of cosmological perturbations evolve independently in linear perturbation theory. Hence, we may always assume that the temperature anisotropies (of primordial origin) observed at point $\vec{x}$ in the direction $\hat{n}$ are given by
\begin{equation}\label{eq:T anisotropies}
	\delta{\hat{T}}(\vec{x},\hat{n})=\int \frac{d^3 k}{(2\pi)^3}\mathcal{T}(\vec{k},\hat{n}) \hat{\phi}(\vec{k}) e^{i \vec{k} \cdot \vec{x}},
\end{equation}
where $\hat{\phi}(\vec{k})$ are the Fourier components of a random field  at a sufficiently early  time, and $\mathcal{T}$ is a transfer function whose explicit form we shall not need.  Say, in a standard $\Lambda$CDM cosmology we have $\hat{\phi}=\hat{\Phi}$, where $\hat{\Phi}(\vec{x})$ is the primordial Newtonian potential, which, because of the absence of anisotropic stress, also equals $\hat{\Psi}(\vec{x})$. Due to the linear relation between temperature anisotropies and primordial perturbations, it immediately follows that zero mean in the primordial perturbations implies zero mean of the temperature anisotropies.

For many purposes, it is more convenient to represent functions on a sphere, like the temperature fluctuations, by their spherical harmonic coefficients
\begin{equation}\label{eq:lm}
	f_{\ell m}=\int d^2 \hat{n} \, f(\hat{n}) \, Y_{\ell m}(\hat{n}).
\end{equation}
Throughout this article we work with \emph{real} spherical harmonics $Y_{\ell m}$, whose properties are summarized in appendix A. To calculate the spherical harmonic coefficients of the temperature anisotropies $a_{\ell m}\equiv \delta T_{\ell m}$ we note that because linear perturbations evolve in an isotropic background (by definition), the transfer function $\mathcal{T}(\vec{k},\hat{n})$ can only depend on the two scalars $k$ and $\hat{k}\cdot \hat{n}$. Hence, we may expand the latter in Legendre polynomials $P_\ell$,
\begin{equation}\label{eq:Legendre expansion}
	\mathcal{T}(\vec{k},\hat{n})=\mathcal{T}(k,\hat{k}\cdot \hat{n})=
	\sum_\ell (2\ell+1)(-i)^\ell \mathcal{T}_\ell (k) P_\ell(\hat{k}\cdot  \hat{n}),
\end{equation}
with real functions $\mathcal{T}_\ell(k)$.  Substituting then equation (\ref{eq:Legendre expansion}) into (\ref{eq:T anisotropies}), setting $\vec{x}=0$,  and using the addition theorem for (real) spherical harmonics in equation (\ref{eq:addition theorem})  we get
\begin{equation}\label{eq:a}
\hat{a}_{\ell m}=4\pi (-i)^\ell \int \frac{d^3 k}{(2\pi)^3} \mathcal{T}_\ell(k) \hat{\phi}(\vec{k}) Y_{\ell m}(\hat{k}).
\end{equation}
Clearly, if primordial perturbations have zero mean, so do the spherical harmonic coefficients of the temperature anisotropies: 
\begin{equation}\label{eq:zero mean}
	\langle \hat{\phi}\rangle=0 \Rightarrow \langle \hat{a}_{\ell m}\rangle =0.
\end{equation}
In particular,  a violation of the condition $\langle \hat{a}_{\ell m}\rangle =0$ would thus imply a violation of statistical homogeneity.

Later we shall also need to know the covariance of the temperature multipoles, which  follows from equation (\ref{eq:a}). If the random field $\hat{\phi}$ is homogeneous and isotropic in the variance, the covariance matrix of the multipoles has elements 
 \begin{equation}\label{eq:hom and iso variance}
 	\langle \Delta \hat{a}_{\ell_1 m_1} \Delta \hat{a}_{\ell_2 m_2}\rangle= C_{\ell_1} \,  \delta_{\ell_1 \ell_2} \delta_{m_1 m_2},
 	\quad \text{with} \quad 
	C_\ell=4\pi\int \frac{dk}{k} \mathcal{P}_\phi(k) \mathcal{T}_\ell(k)^2.
 \end{equation}
 Recall that for arbitrary  $f$ we define
 \begin{equation}
	 \Delta \hat{f}_{\ell m}\equiv \hat{f}_{\ell m}-\langle \hat{f}_{\ell m}\rangle.
 \end{equation}
 Thus, $\delta$ denotes departures from a homogeneous and isotropic background, whereas $\Delta$ denotes deviations from the ensemble mean. In the following, we drop the hat from random variables and fields. 
 
 \subsection{Foreground, Noise and Masks}

Unfortunately, the temperature anisotropies we actually observe in the sky  are not entirely of primordial origin. They are a superposition of primordial anisotropies $\delta T$ and foreground contributions, such as dust emission and synchrotron radiation from our very own galaxy. Appropriate foreground templates allow the WMAP team to eliminate  foregrounds in some regions of the sky \cite{Gold:2010fm}, but the cleaning procedure does not completely remove foreground contamination along the galactic disc. It is thus necessary to subject these maps to additional processing.

Since the actual temperature measurements involve a convolution  with the detector beam $B$, and also include detector noise $N$, we model the temperature anisotropies in a smoothed, foreground-reduced map by
\begin{equation}\label{eq:observed sky}
	\delta T_\text{map}=K*\left[B*(\delta T+F)+N\right],
\end{equation}
where $K$ is the smoothing kernel, the star denotes convolution, and $F$ represents the residual foreground contamination. We assume that the smoothing kernel and the detector beam are rotationally symmetric. This is actually not the case for the WMAP beam, but it should be a good approximation at the scales we are going to consider. In that case, in harmonic space, the convolution acts on the spherical harmonic coefficients simply by multiplication, say,
\begin{equation}
	(B*f)_{\ell m}=B_\ell \, f_{\ell m}.
\end{equation}

In order to remove the residual foregrounds $F$, the contaminated sky regions have to be masked out. Let $M(\hat{n})$ be the corresponding mask, which is defined by ${M(\hat{n}) [K*B*F](\hat{n})=0}$.  Then, by construction, the masked sky $\delta T_M$ does not contain foregrounds,
\begin{equation}\label{eq:masked sky}
	\delta T_M(\hat{n}) \equiv M(\hat{n})\delta T_\text{map}(\hat{n})=M(\hat{n})\ \left[K*B*\delta T+K*N\right](\hat{n}).
\end{equation}
It proves then useful to define an hypothetical ``unmasked" smoothed sky map which is free of foregrounds, but contains the effects of noise and detector beam, and whose multipoles are hence given by
\begin{equation}\label{eq:b}
b_{\ell m}\equiv K_\ell B_\ell a_{\ell m}+K_\ell N_{\ell m}.
\end{equation}

The multipole coefficients of the masked sky $\delta T_M$ are obtained by multiplication with an appropriate convolution matrix. If $M_{\ell m}$ denotes the (real) spherical harmonic coefficients of the mask, it is easy to show that the elements of the convolution matrix are given by 
\begin{equation}\label{eq:mask matrix}
	M_{\ell_1 m_1,\ell_2 m_2}=
	\sum_{\ell m}\frac{1}{\sqrt{4\pi}}D(\ell_1, m_1; \ell, m; \ell_2, m_2)  M_{\ell m},
\end{equation}
where  $D$ is defined in equation (\ref{eq:D}). In particular, the spherical harmonic coefficients of the masked sky are given by
\begin{equation}\label{eq:c}
	c_{\ell m}\equiv (\delta T_M)_{\ell m}=\sum_{\bar{\ell} \bar{m}}M_{\ell m, \bar{\ell} \bar{m}}
	 \, b_{\bar{\ell} \bar{m}}.
\end{equation} 
In practice, we need to work with finite matrices, so we restrict our attention to a finite range of multipole values, $0\leq \ell\leq \ell_\text{max}$. In particular we assume that $M$ is a square matrix.

\begin{figure}
\includegraphics[height=8cm]{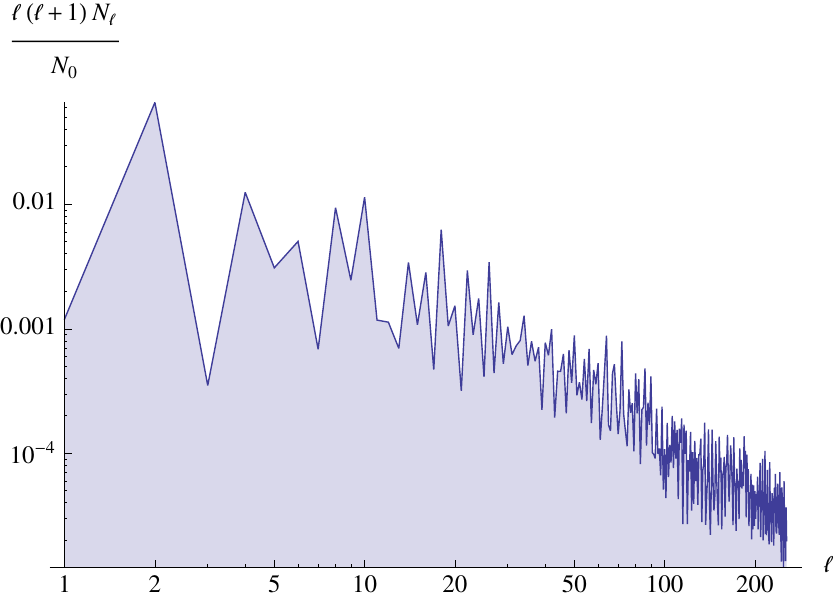}
\caption{The variance of the noise in the temperature anisotropies is inversely proportional to the number of times each sky pixel is observed, $N_\text{obs}(\hat{n})$. The figure shows a plot of the ``power spectrum" $N_\ell\equiv \frac{1}{2\ell+1} \sum_m (N^{-1}_\text{obs})_{\ell m} (N^{-1}_\text{obs})_{\ell m}$ in units of the monopole $N_0$. Weighted by $\ell (\ell+1)$, this captures the degree of anisotropy of the noise variance on angular scales $\sim 180^\circ/\ell$. }\label{fig:noise}
\end{figure}

The WMAP team has found that detector pixel noise is well described by a Gaussian distribution with zero mean \cite{Jarosik:2003fe},  which implies that
\begin{equation}\label{eq:masked mean}
	\langle c_{\ell m} \rangle = \sum_{\bar{\ell} \bar{m}} M_{\ell m, \bar{\ell} \bar{m}}
	 \,   K_{\bar{\ell}} \, B_{\bar{\ell}}\,  \langle a_{\bar{\ell} \bar{m}} \rangle.
\end{equation}
Therefore,  the masked temperature anisotropies have zero mean if the primordial  anisotropies do.  According to the WMAP team \cite{Jarosik:2003fe}, the noise variance  is inversely proportional to the number of times point each pixel is observed.  This is not the same for all pixels, but it is is fairly isotropic (see figure \ref{fig:noise}). Hence, assuming a constant $N_\text{obs}$, and that the noise in different pixels (of area $A$) is uncorrelated, with variance $\sigma_N^2/N_\text{obs}$, we find
\begin{equation}\label{eq:noise covariance}
	\langle  N_{\ell_1 m_1} N_{\ell_1 m_1}\rangle = \mathcal{N}_0 \,
	\delta_{\ell_1 \ell_2} \delta_{m_1 m_2},
\end{equation}
where $ \mathcal{N}_0=A \sigma_N^2/N_\text{obs}$. In any case, at the scales we are interested in, the contribution of the noise to the variance of the masked temperature anisotropies is subdominant. This is however not crucial for our analysis, which simply assumes that the noise satisfies equation (\ref{eq:noise covariance}), without reference to the actual magnitude of $\mathcal{N}_0$.

\subsection{Hypotheses}

In order to test whether the primordial temperature anisotropies have zero mean  (which follows from $\langle \phi \rangle=0$) we need  additional information about the distribution of the harmonic coefficients $a_{\ell m}$. At this point, there is no evidence for non-Gaussian primordial perturbations \cite{Komatsu:2008hk}, so we assume that the latter are normally distributed. In order to uniquely characterize their distribution, it suffices then to  consider their variance.  Among the hypothesis with zero mean in Table \ref{tab:cases}, $H_0$ is the one that underlies most of our analyses of structure. We shall therefore adopt $H_0$ as our null hypothesis.  If $H_0$ is true, then, according to equation (\ref{eq:hom and iso variance}), the variates $\Delta a_{\ell m}= a_{\ell m}$ are independent and have a common variance for the same values of $\ell$. Therefore,  the standard assumption $H_0$ can be cast as a  precise form of the distribution of the temperature multipoles $a_{\ell m}$, which follows from equations (\ref{eq:zero mean}) and (\ref{eq:hom and iso variance}),
\begin{itemize}
	\item[$H_0$:]  Primordial perturbations are normally distributed with zero mean, homogeneous and isotropic variance $\Rightarrow$
	\begin{equation}\label{eq:H0 P}
	dP(a_{\ell m})=\frac{1}{\sqrt{2\pi C_\ell}}
	\exp\left[-\frac{a_{\ell m}^2}{2C_\ell}\right] da_{\ell m}.
	\end{equation}
\end{itemize}
Note that equations (\ref{eq:hom and iso variance}) and (\ref{eq:noise covariance}), together with the definition (\ref{eq:b}),  imply that the covariance matrix of the unmasked, foreground-reduced temperature anisotropies is also diagonal,
\begin{equation}\label{eq:b covariance}
	\langle \Delta b_{\ell_1 m_1} \Delta b_{\ell_2 m_2}\rangle=
	K^2_{\ell_1}(B_{\ell_1}^2 C_{\ell_1}+\mathcal{N}_0) \delta_{\ell_1 \ell_2}\delta_{m_1 m_2}.
\end{equation}

Clearly, if the null hypothesis $H_0$ does not appropriately fit the data, we won't be able to determine whether this is because temperature fluctuations are non-Gaussian, non-isotropic, non-homogeneous, or simply because we used the wrong power spectrum.  We need to analyze  the data in the face of an alternative hypothesis, namely, that primordial perturbations do not have zero mean. Among all the cases  with non-zero mean in table \ref{tab:cases}, the minimal deviation from $H_0$ is hypothesis $H_3$, which also leads to the covariances (\ref{eq:hom and iso variance}). Therefore, we choose  as alternative
\begin{itemize}
	\item[$H_3$:]  Primordial perturbations are normally distributed with non-zero mean  and homogeneous and isotropic variance $\Rightarrow$
	\begin{equation}\label{eq:H3 P}
	dP(a_{\ell m})=\frac{1}{\sqrt{2\pi C_\ell}}
	\exp\left[-\frac{(a_{\ell m}-\langle a_{\ell m}\rangle)^2}{2C_\ell}\right] da_{\ell m}.
	\end{equation}
\end{itemize}
In this case, the covariance of the unmasked, foreground-reduced temperature anisotropies is again given by equation (\ref{eq:b covariance}).

We test  $H_0$ against $H_3$. What singles out our test  is the ability to examine the zero mean hypothesis against the alternative hypothesis of non-zero means.  Indeed, the only difference between $H_0$ and $H_3$ lies in the assumptions about the mean of the perturbations. Without the alternative hypothesis, we would be simply conducting a goodness-of-fit test.

Mathematically it is certainly sensible to postulate hypothesis $H_3$, but the reader may wonder whether $H_3$ is also physically reasonable. In fact, we think it is. Suppose for instance that primordial perturbations are created during an inflationary period in a slightly inhomogeneous universe (after all, if inflation is supposed to explain cosmic homogeneity, it should start  with an inhomogeneous universe.) If we regard these small inhomogeneities as first order perturbations, in linear perturbation theory the properties of the created perturbations---vacuum fluctuations of an appropriate field---only depend on the homogeneous and isotropic background. Hence, the resulting seeded perturbations turn out to be homogeneous and isotropic in the mean and the variance, as in the conventional case, but they have to be added on top of the already existing initial inhomogeneities. In fact, similar ideas have been already discussed in the literature \cite{Donoghue:2007ze,Erickcek:2008sm, Carroll:2008br}.

\section{Test Statistic}

If we just happened to know the temperature multipoles $a_{\ell m}$ (or their foreground-cleaned counterparts $b_{\ell m}$), a test of hypothesis $H_0$ against its alternative $H_3$ would be straight-forward. Under the null hypothesis, for fixed $\ell$, the variables $a_{\ell m}$ form a set of independent, normally distributed variates with zero mean and common variance $C_\ell$. The standard and time-honored way to test the latter involves Student's $t$ statistic,
\begin{equation}
	t\equiv\frac{\sqrt{2\ell+1}\,  \bar{a}_{\ell}}{s_\ell},
\end{equation}
where
\begin{equation}
	\bar{a}_\ell\equiv\frac{1}{2\ell+1}\sum_m a_{\ell m} 
	\quad \text{and}\quad
	s_\ell^2\equiv\frac{1}{2\ell}\sum_m (a_{\ell m}-\bar{a}_\ell)^2
\end{equation}
are, respectively, unbiased estimators of the mean and variance of the distribution. Under the null hypothesis, $t$ follows Student's distribution with $\nu=2\ell$ degrees of freedom, while under the alternative hypothesis, its square is distributed like a ratio of non-central chi-squares (more about this below.) Note that we do not need to make any assumption about the actual values of the $C_\ell$ in order to know how $t$ is distributed under the null hypothesis. 

As we mentioned above, though, it is not possible to subtract part of the galactic contamination, so we are forced to work with the masked sky in equation (\ref{eq:masked sky}). While masking preserves the property of zero mean, equation (\ref{eq:masked mean}), it does not preserve the diagonal form of the covariance matrix. 
Hence, one of the key assumptions of Student's test is lost.  To bring the problem back to the realm of Student's $t$, we shall impose additional symmetries on the problem.

On large angular scales, the main source of foreground contamination stems from the galactic disc, which  can be covered by a mask that spans galactic latitudes in the range ${|b|\lesssim 15^\circ}$. Let us hence assume that the mask is symmetric under rotations around the galactic $z$-axis. Under such rotations, the real spherical harmonics transform according to  equation (\ref{eq:M rotation}). Hence, rotational invariance implies  $M_{\ell m}=0$ for $m\neq 0$. When we substitute the last relation into equation (\ref{eq:mask matrix}) we find, using the results in the appendix,
\begin{equation}
	M_{\ell_1 m_1, \ell_2 m_2}\propto \delta_{m_1 m_2}\quad \text{and} \quad
	M_{\ell_1 m, \ell_2 m}=M_{\ell_1 -m, \ell_2 -m}.
\end{equation}
In other words, the mask matrix is diagonal in $m$ space, and we can basically restrict our attention to $m\geq 0$. Of course, a mask with these symmetry properties  cannot  eliminate all sources of foreground contamination (primarily point sources). We shall address this problem by restricting our analysis to large angular scales, for which the contribution of point sources is small. 

For notational simplicity let us call $M_{(m)}$ the (symmetric) matrix with elements $M_{\ell_1 m, \ell_2 m}$, and let us collect all the multipoles $f_{\ell m}$  for fixed $m$  into a single vector
\begin{equation}
	\vec{f}_m\equiv \left( f_{0m}, f_{1m},\ldots, f_{\ell_\text{max} m}\right),
	\quad \text{with}\,\, f_{\ell m}\equiv 0\,\,  \text{for}\,\, \ell<|m|.
\end{equation}
In this notation then, $M_{(m)}=M_{(-m)}$, and equation (\ref{eq:c}) reads 
\begin{equation}\label{eq:c matrix}
	\vec{c}_m=M_{(m)} \vec{b}_m.
\end{equation}
Suppose now that we find a positive integer $m_\text{max}\leq \ell_\text{max}$ and a vector $\vec{v}$ such that 
\begin{equation}\label{eq:S}
	M_{(m)}\vec{v}=\vec{v}, \quad m=-m_\text{max},\ldots, m_\text{max}.
\end{equation}
In other words, suppose that we find a sky  $v_{\ell m}$ with no components along the contaminated region, such that $v_{\ell m}=v_\ell$ for \emph{all} values of $m$ between $-m_\text{max}$ and $m_\text{max}$. If such a vector exists, it must clearly have vanishing components for $\ell< m_\text{max}$, since $|m|$ cannot exceed $\ell$.
Under these conditions then, the $2m_\text{max}+1$ variables $d_{m}\equiv \vec{v} \cdot \vec{c}_{m}$ (with $|m|\leq m_\text{max})$ satisfy
\begin{equation}\label{eq:e and b}
	d_{m}= \vec{v}\cdot \vec{b}_{m},
\end{equation}
where we have used equation (\ref{eq:S}) and $M_{(m)}=M_{(m)}^T$.  In particular, these variates do not contain galactic residual backgrounds, because they can be constructed from a masked sky, they have zero mean under the null hypothesis,
\begin{equation}
	\langle d_m \rangle=\sum_{\ell=m_\text{max}}^{\ell_{max}}v_\ell \langle b_{\ell m} \rangle,
\end{equation}
and, because of equation (\ref{eq:b covariance}), they have covariances
\begin{equation}\label{eq:covariances}
	\langle \Delta d_{m_1} \Delta d_{m_2} \rangle = \sigma^2 \,
	\delta_{m_1 m_2} 
	 \quad  (|m_1| \leq m_\text{max}, |m_2|\leq m_\text{max}),
\end{equation}
where
\begin{equation}\label{eq:sigma}
	\sigma^2\equiv \sum_{\ell=m_\text{max}}^{\ell_\mathrm{max}}  K_\ell^2 (B_\ell^2 C_\ell+N_0)v^2_\ell. 
\end{equation}
The off-shot of this construction is not only that  the variables $d_{m}$ are uncorrelated (and hence independent),  but also that their variance, equation (\ref{eq:sigma}), is the same for all of them. Note that, by construction,  the variables $\vec{d}_m$ only contain temperature multipoles $b_{\ell m}$ in the range $\ell_\text{min}\leq \ell \leq \ell_\text{max}$, where $\ell_\text{min}=m_\text{max}$.

\subsection{Test Statistic}
Equation (\ref{eq:covariances}) states that the $2m_\text{max}+1$ variates $d_{m}$ ($|m|\leq m_\text{max}$)  form a set of normally distributed \emph{independent} variables, with common variance $\sigma^2$. They are thus statistically analogous to the primordial temperature multipoles  $a_{\ell m}$, which are independent and have the same variance for fixed $\ell$. Under the null hypothesis $H_0$ the $d_{m}$ have zero mean, and under the alternative hypothesis $H_3$ their mean is generically non-zero. Because the $d_{m}$ are linear combinations of the normally distributed $a_{\ell m}$, the distribution of the former is also Gaussian. It is hence natural to choose Student's $t$ as test statistic, although, for later convenience, we shall actually work with its square,
\begin{equation}\label{eq:t squared}
	t^2\equiv \frac{(2m_\text{max}+1)\bar{d}^2}{s^2},
\end{equation}
where 
\begin{equation}\label{eq:t expl}
	\bar{d}\equiv\frac{1}{2m_\text{max}+1}\sum_{|m|\leq m_\text{max}} d_{m}, \quad
	s^2\equiv \frac{1}{2m_\text{max}}\sum_{|m|\leq m_\text{max}} (d_{m}-\bar{d})^2.
\end{equation}
Intuitively, the nature of the test statistic is clear: Up to factors that involve $m_\text{max}$, $t^2$  is just the square of the ratio of sample mean to sample standard deviation. We would expect this ratio to be small if the variables indeed have zero mean, and large if they don't. 

One of the keys of our test statistic is that we know its distribution both under the null and the alternative hypothesis. The identity $\sum_{m} (d_{m})^2=2m_\text{max} s^2+(2m_\text{max}+1) \bar{d}^2$ implies by an extension of Cochran's theorem (section 15.20 in \cite{Kendall&StuartA} and section 35.7 in \cite{Kendall&StuartC}) that the numerator and denominator of equation (\ref{eq:t squared}) are independent variables, both under $H_0$ and $H_3$. Again by the same extension of Cochran's theorem, the numerator (divided by $\sigma^2$)  follows a non-central chi-square  distribution with $\nu_1=1$ degrees of freedom, and non-central parameter 
\begin{equation}
	\lambda_1=(2m_\text{max}+1) \frac{\langle \bar{d}\rangle^2}{\sigma^2},
\end{equation} 
whereas $2m_\text{max} s^2$ (divided by $\sigma^2$) follows a non-central chi-square distribution with ${\nu_2=2m_\text{max}}$ degrees of freedom,  and non-central parameter
\begin{equation}
	\lambda_2=\sum_{|m|\leq m_\text{max}}\frac{(\langle d_{m}\rangle-
	\langle \bar{d}\rangle)^2}{\sigma^2} .
\end{equation}
The statistic $t^2$ is thus a ratio of non-central chi-squares divided by their respective number of degrees of freedom. The latter follows a doubly non-central $F$ distribution, with probability density given by  (section 24.30 in \cite{Kendall&StuartB})
\begin{equation}\label{eq:dP}
	dP(t^2)=e^{-\frac{1}{2}(\lambda_1+\lambda_2)}
	\sum_{r=0}^\infty	\sum_{s=0}^\infty
	\frac{1}{r!s!}\left(\frac{\lambda_1}{2}\right)^r
	\left(\frac{\lambda_2}{2}\right)^s  
	\frac{\nu_1^{\frac{\nu_1}{2}+r}\nu_2^{\frac{\nu_2}{2}+s} (t^2)^{\frac{\nu_1}{2}+r-1}}
	{\left(\nu_2+\nu_1 t^2\right)^{\frac{\nu_1+\nu_2}{2}+r+s}}
	\frac{dt^2}{B(\frac{\nu_1}{2}+r,\frac{\nu_2}{2}+s)},
\end{equation}
where $B$ is the beta function. If all the ensemble means are equal, $\mu\equiv \langle d_{m}\rangle$,  $\lambda_2$ vanishes, and the distribution of $t^2$ simplifies to a non-central $F$-distribution with degrees of freedom $\nu_1=1$ and ${\nu_2=2m_\text{max}}$, and non-central parameter $\lambda\equiv\lambda_1$. Under the null hypothesis $H_0$ the non-central parameter is $\lambda=0$, and the distribution reduces to a central $F$ with $\nu=1$ and $\nu_2=2m_\text{max}$, which is just the square of Student's $t$ distribution. 

We carry out a one-sided test of the null-hypothesis $H_0$ at significance $\alpha$  (say, $\alpha=5\%$) by rejecting the null hypothesis if $t^2$ is larger than $(t^2)_\alpha$, where $(t^2)_\alpha$ is the $\alpha$-point of the central $F(\nu_1=1,\nu_2=2m_\text{max})$ distribution,
\begin{equation}
	 \alpha=P(t^2\geq  (t^2)_\alpha | H_0).
\end{equation}
This amounts to a two-sided test of $H_0$ using Student's $t$, in which we reject the null hypothesis for sufficiently large deviations (of either sign) of $t$ from zero. In order to evaluate the power of this test, $1-\beta$, we need to determine the probability  of rejecting the null hypothesis when the alternative hypothesis $H_3$ is true,
\begin{equation}
	1-\beta\equiv P(t^2\geq(t^2)_\alpha | H_3).
\end{equation}

\begin{figure}
\includegraphics[height=8cm]{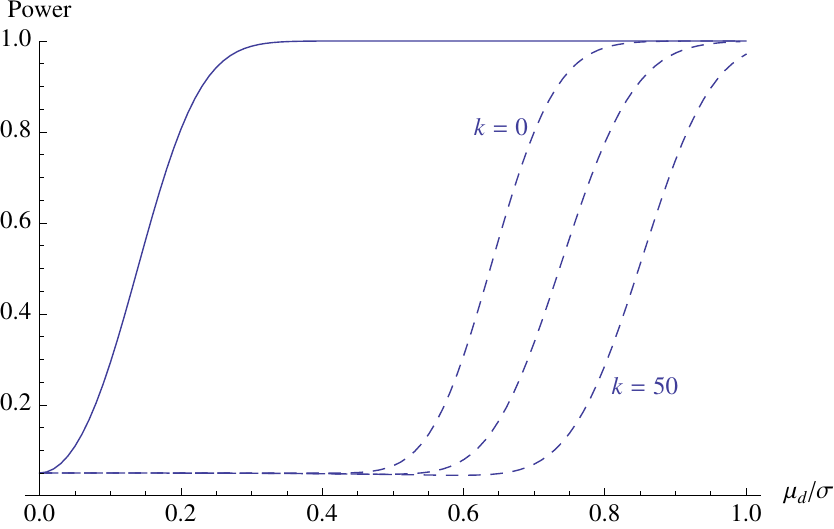}
\caption{Power curves as for a test with  $\alpha=5\%$ significance and $m_\text{max}=100$ as a function of a standardized common mean. The solid curve shows the power of a test based on Student's $t$,  equation (\ref{eq:t squared}). The dashed curves show the power of a test based on the statistic (\ref{eq:tk}) for $k=0, 25$ and $50$.  Clearly, Student's $t$-test is more powerful than any of the $t_k$ tests. (Note that we can restrict our attention to $k\leq 50$,  since for higher $k$ values the statistic $t_k$ has the same distribution as $1/t_{\tilde{k}}$, with $\tilde{k}<50$.) For different choices of parameters, the power curves have the same qualitative form. Evidently, as the number of degrees of freedom $\nu_2=2m_\text{max}$ increases, so does the power of any of these tests.}\label{fig:power}
\end{figure}

The power is a function of the hypothetical standardized non-zero ensemble means $\langle d_{m}\rangle/\sigma$. In the absence of any particular model,  and for the purpose of illustration, we shall consider the simple (though perhaps somewhat unrealistic) case of equals means: $\mu_a\equiv \langle a_{\ell m}\rangle$. Using equations (\ref{eq:e and b}) and (\ref{eq:b}) this translates into
\begin{equation}\label{eq:Z}
	 \mu_d \equiv \langle d_m \rangle = Z \mu_a, \quad
	\text{where} \quad Z\equiv \sum_{\ell=m_\text{max}}^{\ell_{max}} K_\ell B_\ell\,  v_\ell.
\end{equation}
In general, we may regard $\mu_a$  as a measure of the order of magnitude of the mean of the primordial temperature anisotropies in the corresponding multipole range, even if the means are not common.  We plot the power of the test as a function of $\mu_d/\sigma$, for $m_\text{max}=100$,  in figure \ref{fig:power}. We see for instance that in a test with $m_\text{max}=100$, if the mean of our variables is about $0.3$ times their standard deviation (or larger), the $t^2$ statistic will fall in the critical region almost with certainty. 
 
\subsection{Confidence Intervals}
The value of the $t^2$ statistic  tells us not only whether the null hypothesis holds, but also how far from zero the mean may be if the alternative hypothesis $H_3$ is the correct one.  Let  $t^2_{\alpha}$ denote the $\alpha$-point of Student's $t^2$ distribution, and, for simplicity, let us assume again that all the means of the primordial anisotropies have a common value $\mu_a$. Then, since the statistic
\begin{equation}
	t^2_\mu\equiv \frac{(2m_\text{max}+1) (\bar{d}-\mu_d)^2}{s^2}
\end{equation}
is distributed like $t^2$ under the alternative hypothesis $H_3$, we can write $P(t_\mu^2 \leq t^2_\alpha)=1-\alpha$. Following the conventional approach to classical interval estimation, we can cast the last relation as a confidence interval that formally involves $\mu_d$,
\begin{equation}
	P\left(\bar{d}-\frac{t_\alpha\,  s}{\sqrt{2m_\text{max}+1}}\leq \mu_d \leq \bar{d}+\frac{t_\alpha\,  s}{\sqrt{2m_\text{max}+1}}\right)=1-\alpha.
\end{equation}
This is a frequentist interval: If we repeat the same procedure to derive the confidence interval $N$ times, as $N$ approaches infinity our interval will contain  the true mean $\mu_a$ in $(1-\alpha)N$ cases.

\subsection{Extension to Several Statistics}
\label{sec:Extension}

We can also extend this analysis to a set of several independent $t^2$ statistics, designed to probe the temperature anisotropies at different angular scales.  Suppose for instance that we find not one, but a set of $n$ different vectors $\vec{v}{}^{(i)}$ ($i=1,\ldots n$) that satisfy the set of equations (\ref{eq:S}), with  $m_\text{max}=m_\text{max}^{(i)}$ and $\ell_\text{max}=\ell_\text{max}^{(i)}$. Then, for fixed $i$, the variates $d_m^{(i)}\equiv \vec{v}^{(i)}\cdot \vec{c}_m$   still have a common variance, as in equation (\ref{eq:covariances}). We can therefore define a set of $n$ different  statistics $t^2$ simply by replacing $\vec{d}$ by $\vec{d}{}^{(i)}$ and $m_\text{max}$ by $m_\text{max}^{(i)}$ in equations (\ref{eq:t squared}) and (\ref{eq:t expl}).  But if the ranges $m_\text{max}^{(i)} \leq \ell \leq \ell_\text{max}^{(i)}$ are disjoint,  the vectors $\vec{v}^{(i)}$ do not have any common element, and the variables $\vec{d}{}^{(i)}$ and  $\vec{d}{}^{(j)}$ are also  uncorrelated for $i\neq j$.  In that case, the different $t^2$ statistics  are mutually independent, and probe the temperature anisotropies in the disjoint multipole ranges 
\begin{equation}\label{eq:interval}
	\ell_\text{min}^{(i)} \leq \ell \leq \ell_\text{max}^{(i)}, \quad \text{with}
	\quad \ell_\text{min}^{(i)}=m_\text{max}^{(i)}.
\end{equation}

\subsection{Alternative Tests}
\label{sec:alternative statistics}
 
A perhaps undesirable property of our test statistic is that $t^2$ is not a scalar under rotations. Since the mask breaks rotational symmetry anyway, this is not a problem by itself. Nevertheless, the mask does preserve the symmetry under azimuthal rotations, so it would be natural to demand at least invariance of our statistic under this unbroken subgroup. 

Under an azimuthal rotation by an angle $\phi$, the variates $d_m$ transform as in equation (\ref{eq:M rotation}), with $a_{\ell m}$ replaced by $d_m$.  Clearly, the statistic  $t^2$ in equation (\ref{eq:t squared}) does not remain invariant under such rotations. But the transformation law (\ref{eq:M rotation}) immediately suggests how to address the problem. Indeed, the variables
\begin{equation}
	q_0^2\equiv d_0^2, \quad q_m^2\equiv \frac{1}{2}\left(d_m^2+d_{-m}^2\right)\quad (m>0)
\end{equation}
are invariant under azimuthal rotations by construction, and they also share the same variance. Hence, any ratio of the sum of two disjoint subsets of these squared variables is distributed like a ratio of independent (eventually non-central)  $\chi^2$ distributions. 

Consider for instance the set of statistics
\begin{equation}\label{eq:tk}
	t^2_k \equiv  \frac{m_\text{max}-k}{1+k} \frac{\sum\limits_{m=0}^k q_m^2}{\sum\limits_{m=k+1}^{m_\text{max}} q_m^2},
\end{equation}
where $k$ is an arbitrary parameter. As before, under the alternative hypothesis $H_3$ the probability density of $t^2_k$  is given by equation (\ref{eq:dP}), with degrees of freedom and non-central parameters given by, respectively,
\begin{align}
	\nu_1&=1+k, &
	 \lambda_1 \sigma^2 &=\langle d_0 \rangle^2
	+\frac{1}{2}\sum\limits_{m=1}^k \left(\langle d_m\rangle^2+\langle d_{-m}\rangle^2\right), \\
	\nu_2 &=m_\text{max}-k, & 
	\lambda_2 \sigma^2 &=\frac{1}{2}\sum\limits_{m=k+1}^{m_\text{max}}
	\left(\langle d_m\rangle^2+\langle d_{-m}\rangle^2\right).
\end{align}
In particular, under the null hypothesis $H_0$, $t_k^2$ follows a central $F$ distribution with $\nu_1$ and $\nu_2$ degrees of freedom.

Using equation (\ref{eq:dP}) we calculate the power of the set of alternative tests based on the statistic (\ref{eq:tk}). Assuming that all the means are common, as above, we find the power curves in figure \ref{fig:power}. Inspection of the figure quickly reveals that the test  based of Student's $t^2$ statistic is uniformly more powerful than any test based on a $t^2_k$ statistic. This is no coincidence at all; Student's $t$ test is widely employed because of its optimal properties (see example 23.14 in \cite{Kendall&StuartB}.) Therefore, in this article we just focus on Student's $t^2$. As long as we stick to a single (random) orientation of the sky, our results have a straight-forward statistical interpretation, since all we need to know is how the test statistic is distributed for an arbitrary (but fixed) sky orientation. In addition, the non-scalar nature of the statistic may help us to identify that area of the sky eventually responsible for a violation of the zero mean hypothesis. To  conclude, we should also point out that Student's $t$ test is known to be robust to departures from normality, at least for \emph{independent} variables drawn from the same distribution (section 31.3 in \cite{Kendall&StuartB}).

\section{Data and Analysis}

Our data analysis pipeline consists of four main steps. First, we construct an appropriate mask to eliminate residual galactic foregrounds. Then, we identify a vector $\vec{v}$ that belongs to the range of all the mask matrices $M_{m}$ for $|m|\leq m_\text{max}$. We degrade the  cosmic microwave maps to lower resolution, and apply our test statistic to these maps. Finally, we check for an eventual residual contamination in our results. This section lists the details of each of these steps.  The reader not interested in technical details is welcome to skip this part and jump to the next section for the actual results.

\subsection*{Mask}
As mentioned above, in order to eliminate  galactic contamination, and preserve azimuthal symmetry at the same time, we construct a mask invariant under rotations along the galactic $z$-axis. Our starting point is a HEALPix\footnote{{\tt http://healpix.jpl.nasa.gov/}} pixelization of the sphere with ${N_\text{side}=64}$.  We set all pixels in the mask with  galactic latitude $|b|\le 20 ^\circ$ to zero, and all the remaining pixels  to one. The effective area covered by the mask is $66\%$ of the full sky. We label the components of the pixelized mask by $M(\hat{n}_i)$, where $i$ runs over all the $N_\text{pix}\equiv 12\times 64^2=49152$ pixels of the mask. This particular mask is in fact also symmetric under parity, but we do not make explicit use of this symmetry.

\subsection*{Vector $\vec{v}$}

In order to find a $\vec{v}$ that satisfies equation (\ref{eq:S}), we look for a common solution of the set of equations
 \begin{equation}\label{eq:null}
 	({\mathbb 1}-M_{(m)})\vec{v}=0,  \quad m=-m_\text{max},\ldots, m_\text{max}.
 \end{equation}
Since $M_{(m)}$ is the mask matrix, the linear operator ${\mathbb 1}-M_{(m)}$  gives the components of the vector along the contaminated galactic region. Hence, equation (\ref{eq:null}) states that the vector $\vec{v}$ should have a vanishing component along such region. 

In order to find the components of $\vec{v}$, it is numerically more convenient to work in real space. We fix the values of $\ell_\text{max}$ and  $m_\text{max}$ and calculate a matrix $D$ whose elements are defined by
\begin{equation}
	D_{i \ell}=(1-M(\hat{n}_i)) \sum_{|m| \leq m_\text{max}} Y_{\ell m}(\hat{n}_i),
\end{equation}
where $i$  runs over all pixels in a HEALPix pixelization of the sphere with $N_\text{side}=64$, and $\ell=m_\text{max},\ldots, \ell_\text{max}$.  Then, the set of equations (\ref{eq:null}) reads $\sum_\ell  D_{i\ell}\, v_\ell=0$, or
\begin{equation}\label{eq:Cv}
	 D\, \vec{v}=0.
\end{equation}
We find an approximate solution of  equation (\ref{eq:Cv}) by singular value decomposition,
\begin{equation}
	D=\sum_{\alpha} \vec{U}_\alpha \, \Sigma_\alpha \, \vec{V}^T_\alpha.
\end{equation}
Here, the $\vec{U}_\alpha $ is a set of $N_\text{pix}$-dimensional orthonormal vectors,  the $\Sigma_\alpha$ are the singular values (arranged in order of decreasing magnitude), and  the $\vec{V}$ is a set $(\ell_\text{max}-m_\text{max}+1)$-dimensional orthonormal vectors.  We choose the vector $\vec{v}$ to be the last right singular vector, that is, $\vec{v}=\vec{V}_\alpha$, with $\alpha=\ell_\text{max}-m_\text{max}+1$. We label the corresponding singular eigenvalue $\Sigma_\text{last}$.

\begin{figure}
\includegraphics[height=8cm]{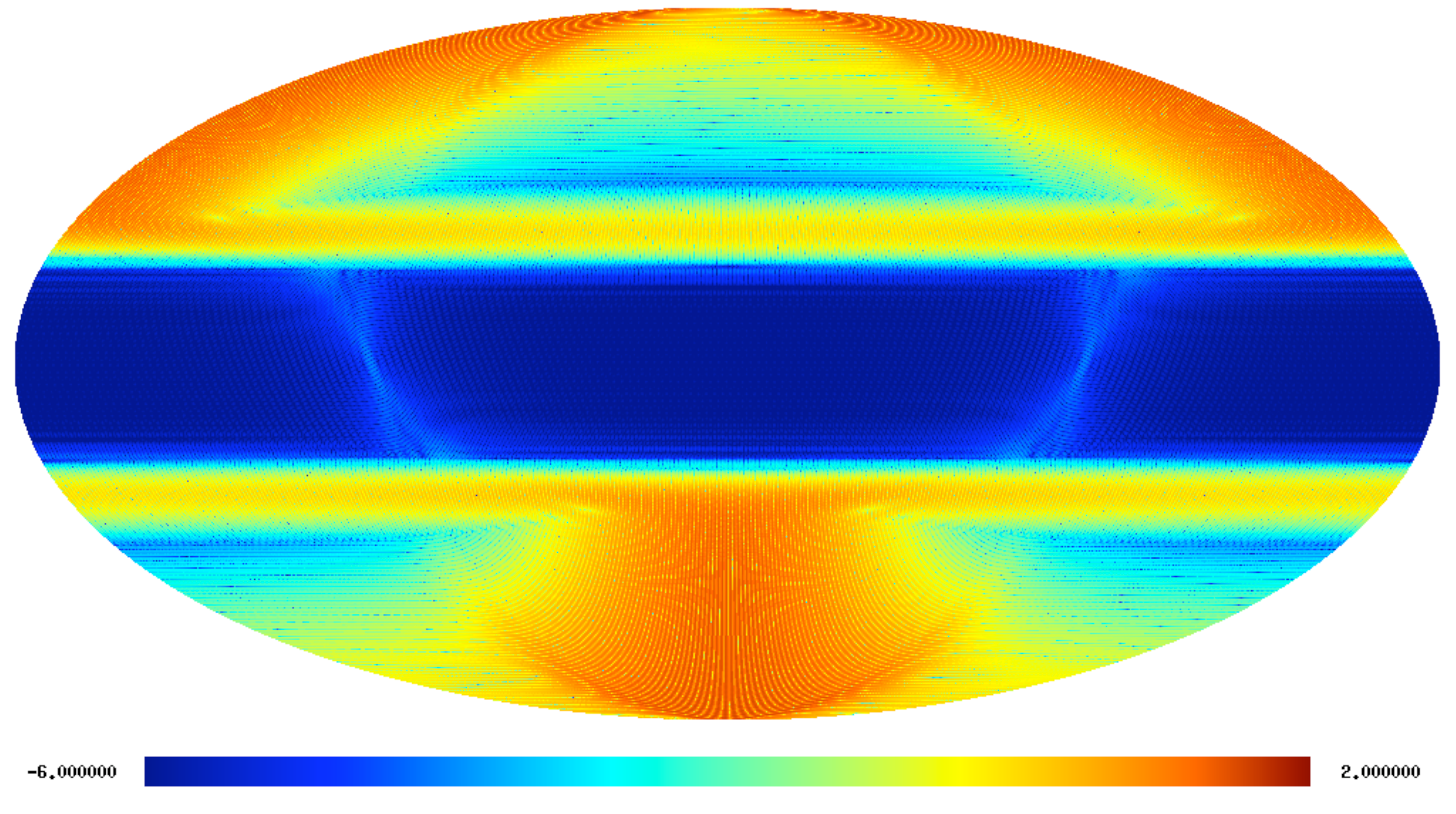}
\caption{The logarithm of the absolute value of the sky defined by equation (\ref{eq:v sky}), for $\ell_\text{max}=212$ and $m_\text{max}=177$. Those regions of the sky with the largest values are those that are more heavily weighted in our statistic, as implied by equation (\ref{eq:e and b}). As seen in the figure, the galactic region is basically excluded from our analysis. For other choices of $\ell_\text{max}$ and $m_\text{max}$ the structure of the sky is the same, as long as the singular eigenvalue $\Sigma_\text{last}$ remains sufficiently small.\label{fig:v}}
\end{figure}

Because the vectors $\vec{U}_\alpha$ and $\vec{V}_\alpha$ are orthogonal, the singular value $\Sigma_\text{last}$  is the norm of  $D\, \vec{v}$,
\begin{equation}
	\sum_i \left(\sum_\ell  D_{i\ell}\, v_\ell\right)^2=\Sigma_\text{last}^2.
\end{equation} 
In general, this singular value is non-zero, so our solution of equation (\ref{eq:Cv}) is not exact but only approximate. Modulo a normalization factor, the value of $\Sigma_\text{last}$ is then an indicator of the potential degree of contamination, i.e.,  the overlap between our vector $\vec{v}$ and the contaminated galactic region. The latter typically increases with increasing $m_\text{max}$, since the number of non-zero elements of $v_\ell$ freely available to solve equation (\ref{eq:Cv}) decreases with increasing $m_\text{max}$. Because the power of the $t^2$ test increases with the number of degrees of freedom, $\nu_2=2m_\text{max}$, we choose the maximum possible value of $m_\text{max}$ for which $\Sigma_\text{last}$, divided by the norm of the mask times the norm of sky encoded in $\vec{v}$, remains under $5\cdot 10^{-8}$.  Indeed, the resulting vector can be represented visually, by defining the sky
\begin{equation}\label{eq:v sky}
	v_{\ell m}=
	\begin{cases} 
		v_\ell, & |m|\leq m_\text{max}, \,  m_\text{max}\leq \ell \leq \ell_\text{max} \\
		0, & \text{otherwise},
	\end{cases}
\end{equation}
which captures those regions of the sky that enter our statistic. As an example, the corresponding real space sky for $\ell_\text{max}=212$ and $m_\text{max}=177$ is shown in figure \ref{fig:v}. 

The same process can be repeated for different choices of $\ell_\text{max}$ and $m_\text{max}$. If the corresponding intervals (\ref{eq:interval}) do not overlap, we can use the resulting set of vectors $\vec{v}$ to construct a set of mutually independent $t^2$ statistics, as explained in subsection \ref{sec:Extension}.

\subsection*{Maps and Processing}

We analyze the seven-year (version 4), full resolution, foreground-reduced  Q2, V2 and W1 differencing assembly cosmic microwave anisotropy maps provided by the WMAP mission.\footnote{Available at \texttt{http://lambda.gsfc.nasa.gov/}.} Since these maps probe different frequencies of the microwave spectrum, a dependence of our results on the particular map would indicate  non-thermal foreground contamination.  We expect the latter to be smallest for the W and V maps, and largest for the Q map. We thus take the W1 assembly to be our fiducial map, and keep the V2 and Q2  assemblies just for comparison. 

The WMAP has subtracted the dipole and primordial monopole from the three differencing assemblies,  and the latter have been  smoothed with a  Gaussian kernel $K$ of $\text{FWHM}=1^\circ$.  The sky maps are expanded into (real) spherical harmonics, and band-limited to a maximum multipole value $\ell=\ell_{\max}$. The different values of $\ell_\text{max}$ are  chosen iteratively to cover the multipole range $0\leq \ell \leq 212$ with non-overlapping intervals.  We choose $\ell=212$ as the absolute maximum for $\ell$ because we expect point sources to significantly contaminate the temperature anisotropies at higher multipoles.  Note that it is not necessary to mask the sky prior to processing, since  the vector $\vec{v}$ has no components along the galactic region by construction. 

\subsection*{Contamination}

There are two main possible sources of systematic errors in our analysis: Galactic contamination due to an insufficiently resolved mask or an imperfect solution of equations (\ref{eq:null}), and point source contamination due to unmasked high-latitude point sources.  In order to estimate both  we basically follow the same approach.

Let us assume that the values of $\ell_\text{max}$ and $m_\text{max}$ have been fixed. To estimate the amount of galactic contamination, we subtract from the cosmic microwave maps the portion of the sky covered by the WMAP seven-year temperature analysis mask and run the resulting sky map through the analysis pipeline described above. The change in the {$P$-value} of the $t^2$ statistic is then a measure of galactic contamination.  For $\ell_\text{max}=212$ and $m_\text{max}=177$ for instance, the change in the $t^2$ statistic after subtraction of the galaxy is less $0.01\%$ for the W1 map.

Similarly, to estimate the amount of  point source contamination we construct a sky map of temperature anisotropies from the point  source fluxes listed in the WMAP point-source catalog \cite{Wright:2008ib}. We subtract the point source map from the cosmic microwave background and run the resulting sky map through our data analysis pipeline. The change in the corresponding {$P$-value} of the $t^2$ statistic is then  a measure of point source contamination. Certainly, there are unresolved point sources that the WMAP catalog does not contain, but the contribution of these sources is small compared to the contribution of the actually detected sources that we are not able to mask. Say, if we choose $\ell_\text{max}=212$ and $m_\text{max}=177$, the change in the $t^2$ statistic after point source subtraction is less $1\%$ for the W1 map. Point sources do not typically have thermal spectra, so an inspection of our results for different differencing assemblies gives us yet another handle on such contamination. 

\begin{figure}
\includegraphics[height=8cm]{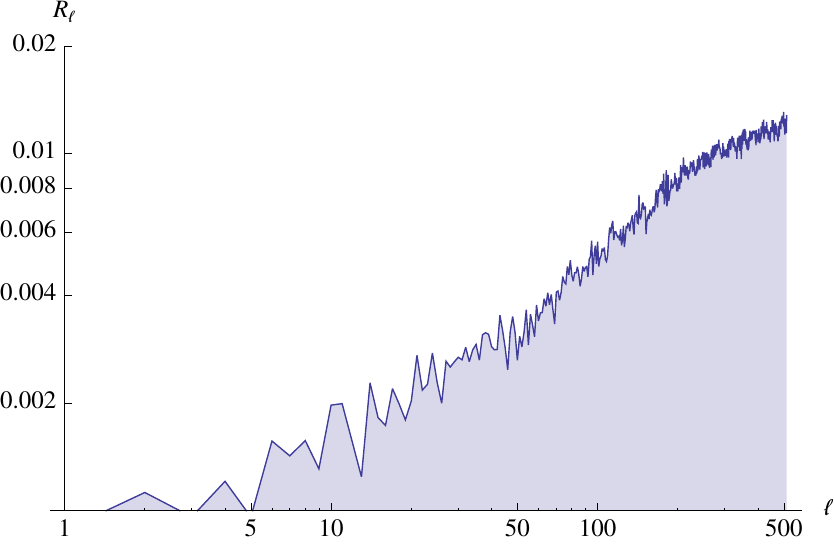}
\caption{Power in detected W-band  point sources relative to actual cosmic microwave anisotropies, equation (\ref{eq:R}). The point source contribution is subdominant at low multipoles, and reaches $1\%$ at $\ell\approx 256$.}
\label{fig:point sources}
\end{figure}

An alternative way to estimate point source contamination involves the ratio
\begin{equation}\label{eq:R}
	R_\ell\equiv \frac{\sum_m s_{\ell m}^2}{\sum_m a_{\ell m}^2},
\end{equation}
where the $s_{\ell m}$ are the multipoles of the temperature map constructed from detected point sources alone, and the $a_{\ell m}$ are the spherical harmonic coefficients of the analyzed sky (say, the $W1$ map).  The sums in $R_\ell$ only run over $m$, because our statistics are sensitive only to a relatively small window of multipoles in $\ell$ space.  As shown in figure \ref{fig:point sources}, $R_\ell$ remains below $1\%$ up to $\ell\approx 256$, which is in broad agreement with our direct estimate of point source contamination using the $t^2$ statistic. Because point source contamination decreases with  the frequency of the map \cite{Wright:2008ib},  the W1 differencing assembly is  less contaminated than the other two assemblies. This is why we take W1 to be our fiducial map.  

Note by the way  that none of the procedures described above is an actual attempt to subtract galactic or point source contamination. Instead, it is just a way to estimate the contribution of the unmasked foregrounds to the $P$-value of our statistic.

\section{Results}

\subsection{$P$-values}

\begin{figure}
\includegraphics[height=8cm]{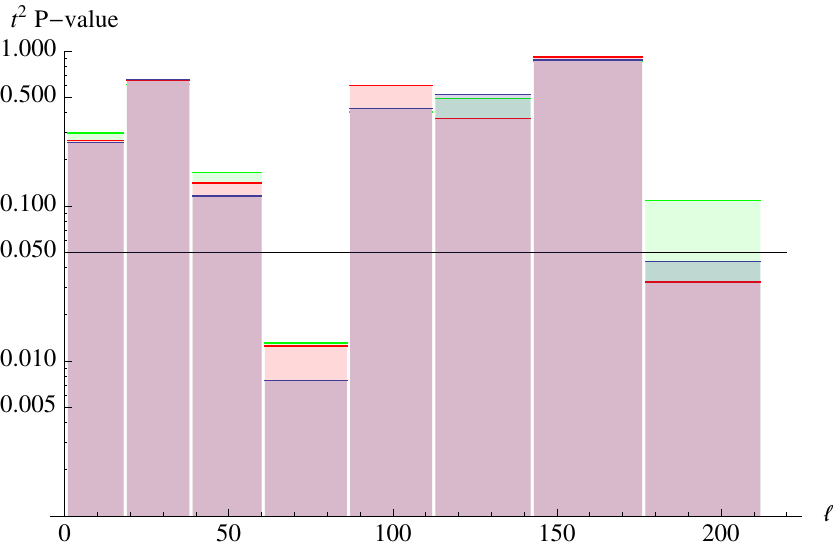}
\caption{Visual representation of table \ref{tab:results}. We plot the $P$-values of the $t^2$ statistic under the null-hypothesis for different multipole ranges and different differencing assemblies (blue for $W1$, red for $V2$, green for $Q2$.)  For reference, the horizontal line marks $5\%$ probability. Clearly, the value of $t^2$ in the multipole range $61\leq \ell \leq 86$ is anomalously small.}
\label{fig:P values}
\end{figure}

Our results are summarized in table \ref{tab:results} and represented graphically in figure \ref{fig:P values}. There are two multipole ranges in which the $P$-value of our $t^2$ statistic is smaller than $5\%$. In the first case, for $\ell=61 \to 86$, the $P$-value across the three differencing assemblies remains under $1.5\%$, so it does not seem that this result is due to residual foregrounds alone. In particular, in this multipole range point source contamination is negligible. In the second case, for $\ell=177 \to 212$, the $P$-value of the statistic for the Q2 map is normal, so we may tentatively attribute the difference in the values of $t^2$ among maps to foreground contamination. This explanation however is somewhat problematic, because we expect foregrounds to make the actual value of $t^2$ less likely, and because contamination is stronger in the Q2 map, whose $t^2$ is normal.

Can we speak then of statistically significant evidence against the null hypothesis? To answer this question, we need to realize that we have constructed a set of $n=8$ independent tests of the null hypothesis, one for each multipole range. Hence, if we would like the \mbox{$P$-values} of all individual statistics to be larger than $\alpha$ with probability $1-\alpha_\text{tot}$ (under the null hypothesis), we should choose the size  $\alpha$ of each individual test to satisfy
\begin{equation}
	1-\alpha_\text{tot}=(1-\alpha)^n.
\end{equation}
For $n=8$ and $1-\alpha_\text{tot}=95\%$, this yields $\alpha=0.64\%$. None of the $P$-values in table \ref{tab:results} is as low.

We reach similar conclusions by calculating the value of a statistic often used to combine the results of multiple independent tests of a single hypothesis: Stouffer's weighted $Z$ test \cite{Cousins}.  Let $P_i$ be the $P$-value of our $t^2$ test in the $i$-th multipole range, and let $Z_i$ be the $P_i$ point of a standard normal distribution.  Then, the variate
\begin{equation}
	Z\equiv \frac{\sum_{i=1}^n w_i Z_i}{\sqrt{\sum_{i=1}^n w_i^2}},
\end{equation}
where the $w_i$ are the weights assigned to each test, follows a normal distribution with zero mean and unit variance. We weigh each test by the number of degrees of freedom of the corresponding $t^2$ test, ${\nu_i=2 m_\text{max}^{(i)}}$. For the eight $P$-values listed in table \ref{tab:results}, and the W1 differencing assembly map, the value of Stouffer's statistic is
\begin{equation}\label{eq:Z}
	Z=1.21, 
\end{equation}
clearly under two sigma away from the mean of the standard normal distribution. 

\begin{table}
$
\begin{array}[t]{ r r  r@{.}l  r@{.}l  r@{.}l  r@{.}l  r@{.}l  r@{.}l  }
 \hline \hline
\multicolumn{2}{c}{\text{Range}}   & \multicolumn{4}{c}{Q2} & \multicolumn{4}{c}{V2} & \multicolumn{4}{c}{W1}
 \\
 \ell_\text{min} & \ell_\text{max} &  \multicolumn{2}{r}{t^2} &  \multicolumn{2}{l}{\text{$P$-value}}  
  &  \multicolumn{2}{r}{t^2} &  \multicolumn{2}{l}{\text{$P$-value}}  
   &  \multicolumn{2}{r}{t^2} &  \multicolumn{2}{l}{\text{$P$-value}}  
 \\
 \hline \hline
 1 &18		& 1&957 & 29&7\% 	& 2&367 & 26&4\%		& 2&457 & 25&8\% \\
 19 & 38		& 0&269 & 60&7\%	& 0&208 & 65&1\% 		& 0&200 & 65&7\% \\
 39 & 60		& 1&963 & 16&5\%	& 2&212 & 14&1\%	 	& 2&525 & 11&6\% \\
 61 & 86 		& 6&341 & 1&3\%	& 6&431 &  1&2\% 		& 7&397 &  0&7\% \\
 87 & 112		& 0&699 & 40&4\%	& 0&275 & 60&1\%		& 0&639 & 42&5\% \\
 113 & 142	& 0&473 & 49&2\%	& 0&816 & 36&7\% 		& 0&406 & 52&5\% \\
 143 & 176	& 0&029 & 86&5\%	& 0&012 & 91&4\%		& 0&025 & 87&5\% \\
 177 & 212	& 2&582 & 10&9\%	& 4&615 &  3&2\% 		& 4&091 &  4&4\% \\
 \hline\hline
\end{array}
$
\caption{The value of the $t^2$ statistic and the corresponding $P$-value (the probability for the statistic to be larger than the value we observe.) The range column indicates the multipoles that enter the calculation of the statistic ($\ell_\text{min}\leq \ell \leq\ell_\text{max}$). We perform the same analysis for three different differential assembly maps, $Q2$, $V2$ and $W1$. The $P$-value is anomalously small (less than $5\%$) for all three maps just in one multipole range: 61 to 86. }
\label{tab:results}
\end{table}

\subsection{Goodness of Fit}

The reader may also wonder how much better the data are fit by a distribution with non-zero mean. In order to find that out, we calculate first an effective $\chi^2$ by extremizing the likelihood under both the null and the alternative hypothesis,
\begin{equation}\label{eq:chisq}
	\chi_i^2\equiv -2 \mathcal{L}_\text{max}(\vec{d}\, | H_i).
\end{equation}
Since the variates $d_m$ are normal and independent, the likelihood is simply a product of  Gaussian density functions.  Therefore, sample mean sample variance respectively are the maximum-likelihood estimators for the population mean and the population variance.

The difference $\Delta\chi\equiv \chi_0^2-\chi_3^2$ is a measure of how much the fit improves when we relax the assumption of zero mean. Because of equation (\ref{eq:chisq}), this difference is a monotonic function of the ratio of maximum likelihoods under $H_0$ and $H_3$, which also happens to be a monotonic function of the $t^2$ statistic (example 24.1 in \cite{Kendall&StuartB}). For illustration, we list the corresponding values of $\Delta\chi^2$ in table \ref{tab:IC}. Clearly, since we have an additional parameter to fit the data, we expect a better fit under $H_3$. To correct for the presence of additional parameters, several model selection measures have been proposed in the literature \cite{Liddle:2007fy}. In table \ref{tab:IC} we list the difference in the corrected Akaike information criterion (AIC$_\text{c}$) and the difference in the Bayesian Information Criterion (BIC). From a Bayesian perspective, the difference in information criteria $\Delta$  is a measure of relatively model likelihood 
\begin{equation}
	\frac{\mathcal{L}(H_0|\vec{d}\,)}{\mathcal{L}(H_3|\vec{d}\,)}=\exp\left(-\frac{\Delta}{2}\right).
\end{equation}
This equation allows us then to interpret $\Delta/2$ as a number of standard deviations. Again, a distribution with non-zero mean seems to be a better model to describe the data in the multipole range $61\to 86$.  But as we emphasized above, this is relatively likely to happen if multiple ranges of multipoles are considered.

\begin{table}
$
\begin{array}[t]{ r r r  r@{.}l  r@{.}l  r@{.}l   }
 \hline \hline
 \ell_\text{min} & \ell_\text{max} & \text{dof}
  &  \multicolumn{2}{c}{\Delta\chi^2} &  \multicolumn{2}{c}{\Delta\text{AIC}_\text{c}}  
   &  \multicolumn{2}{c}{\Delta\text{BIC}} 
 \\
 \hline \hline
 1 &18		&  3	 	& 2&40	& \multicolumn{2}{c}{-}	& 1&31 	\\
 19 & 38		& 39	 	& 0&21  	& -2&02				&  -3&46 	\\
 39 & 60		& 79	 	& 2&52	& 0&41				&  -1&85 	\\
 61 & 86 		& 123 	& 7&24 	& 5&17				&  2&43 	\\
 87 & 112		& 175	& 0&64 	& -1&40				&  -4&52 \\
 113 & 142	& 227	& 0&41	& -1&63				&  -5&02	\\
 143 & 176	& 287	& 0&02	& -2&00				&  -5&63 	\\
 177 & 212	& 355	& 4&08	& 2&06				&  -1&79 	\\
 \hline\hline
\end{array}
$
\caption{Comparison of fits to the W1 data under the null and alternative hypotheses. Positive values indicate that the alternative hypothesis is a better description of the data.  The change in $\chi^2$ simply shows that the temperature anisotropies are better fit in the presence of additional parameter. The AIC$_\text{c}$ corrects for the latter, and indicates strong evidence for the alternative hypothesis in the multipole range $61\to86$. The BIC heavily penalizes the presence of an additional parameter, but still indicates substantial evidence for a non-zero mean  in that multipole range.}
\label{tab:IC}

\end{table}

\subsection{Confidence Intervals}

\begin{figure}
\includegraphics[height=8cm]{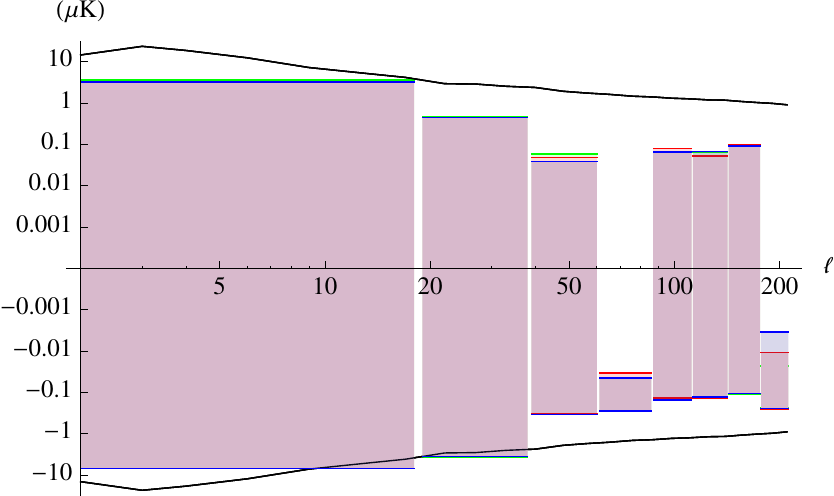}
\caption{Limits on the value of a common mean at $95\%$ confidence limit (in $\mu$K units), as in table \ref{tab:limits}. Again, blue, red and green label the limits derived from the W1, V2 and Q2 band maps respectively. The extent of the interval on the $\ell$ axis indicates the range in values of $\ell$ for which the limit applies. For comparison we also plot the variance of the multipole components for the best WMAP's estimate of the binned  power spectrum, $\pm \sqrt{C_\ell}$. Note that the temperature scale is logarithmic, with positive and negative values on either side of the axis. }
\label{fig:limits}
\end{figure}

The actual values of the test statistic for our choices of $\ell_\text{max}$ and $m_\text{max}$ also allow us to place the first  limits on the magnitude of an eventual common mean of the primordial perturbations in the given range of multipoles.  These limits are collected in table \ref{tab:limits} and graphically represented in figure \ref{fig:limits}.  At angular scales smaller than about four degrees,  the limits are typically one order of magnitude below the standard deviation of the temperature multipoles. In two cases, the confidence interval does not contain zero, which is again an expression of an anomalously high value of the $t^2$ statistic in the corresponding multipole range. But, as before, since these are $95\%$ confidence intervals, the probability that all of them contain the true mean is only $66\%$. In any case, these limits should not be taken too literally. The assumption of a common mean is somewhat unrealistic, so these intervals should be rather interpreted of an order of magnitude estimate of possible deviations from the zero-mean assumption, even if the means of the anisotropies do not share a common value in the corresponding multipole range. 

\begin{table}
$
\begin{array}[t]{ r r  r@{.}l  r@{.}l  r@{.}l  r@{.}l  r@{.}l  r@{.}l  r@{.}l  }
 \hline \hline
\multicolumn{2}{c}{\text{Range}}   & \multicolumn{4}{c}{Q2} & \multicolumn{4}{c}{V2} & \multicolumn{4}{c}{W1} &  \multicolumn{2}{c}{}
 \\
 \ell_\text{min} & \ell_\text{max} &  \multicolumn{2}{r}{\mu_\text{min}} &  \multicolumn{2}{l}{\mu_\text{max}}  
  &  \multicolumn{2}{r}{\mu_\text{min}} &  \multicolumn{2}{l}{\mu_\text{max}}  
   &  \multicolumn{2}{r}{\mu_\text{min}} &  \multicolumn{2}{l}{\mu_\text{max}}  
   & \multicolumn{2}{r}{\sqrt{C_\ell}}
 \\
 \hline \hline
 1 &18		& -7&061  &3&596		& -6&989 & 3&308		& -6&962 & 3&244	& 7&20	\\
 19 & 38		& -0&275  & 0&464		& -0&283 & 0&448	& -0&278 & 0&438	& 2&87	\\
 39 & 60		&  -0&335 & 0&058		& -0&336 & 0&049	& -0&347 &  0&039	& 1&92	\\
 61 & 86 		& -0&278  & -0&033		& -0&278 & -0&034 	& -0&286 & -0&045	& 1&46	\\
 87 & 112		& -0&156  & 0&063		& -0&137 & 0&079	& -0&153 & 0&065	& 1&30	\\
 113 & 142	&  -0&134 & 0&065		& -0&140 & 0&052 		& -0&129 & 0&066	& 1&17	\\
 143 & 176	& -0&112  & 0&094		& -0&106 & 0&095		& -0&108 & 0&092	& 1&09	\\
 177 & 212	& -0&230  & 0&023		& -0&251 & -0&011		& -0&249 & -0&003	& 0&98	\\
 \hline\hline
\end{array}
$
\caption{Lower and upper limits on a common mean of the primordial temperature multipoles $\mu_a$ in the given multipole range at $95\%$ confidence level. Temperature units are $\mu$K. For comparison we list WMAP's best-fit estimate of the binned power spectrum around the center of the corresponding multipole range.}
\label{tab:limits}
\end{table}

\section{Conclusions}

Our results show  significant evidence for a non-zero mean of the temperature multipoles in the range $\ell=61$ to $\ell=86$, at the $99.3\%$ confidence level. Taken as a whole however, because this range is just one among eight different multipole bins, the evidence against the zero-mean assumption is  statistically insignificant, falling under the $95\%$ confidence level.

Whatever the case, the limits we have set on the mean of the primordial anisotropies in a set of multipole bins indicate that an eventual non-zero mean has to be about an order of magnitude smaller than the standard deviation of the temperature anisotropies. In that sense, observations constrain the mean to be small. In retrospective, we have therefore partially  justified the common assumption of vanishing mean of the cosmological perturbations.

\acknowledgments

This work is supported in part by the NSF Grant PHY-0855523. Some of the results in this paper have been derived using the HEALPix \cite{Gorski:2004by} package. We acknowledge the use of the Legacy Archive for Microwave Background Data Analysis (LAMBDA). Support for LAMBDA is provided by the NASA Office of Space Science.

\appendix
\section{Real Spherical Harmonics}
In this article we expand functions defined on a sphere in \emph{real} spherical harmonics $Y_{\ell m}$. These are related to the conventional complex spherical harmonics $\mathcal{Y}_{\ell m}$ by 
\begin{equation}\label{eq:sh}
Y_{\ell m}\equiv 
	\begin{cases}
	 \sqrt{2} \, \text{Im} \, \mathcal{Y}_{\ell -m}, & m<0 \\
	 \mathcal{Y}_{\ell m}, & m=0 \\
	 \sqrt{2} \, \text{Re} \, \mathcal{Y}_{\ell m}, & m>0 .
	\end{cases}
\end{equation}
It follows that the real multipole coefficients $a_{\ell m}$ and their complex counterparts $\mathcal{A}_{\ell m}$ are related to each other by
\begin{equation}\label{eq:relations}
a_{\ell m}=
\begin{cases}
		-\sqrt{2} \, \text{Im} \, \mathcal{A}_{\ell -m}, & m<0 \\
		 \mathcal{A}_{\ell m}, & m=0 \\
		 \sqrt{2} \, \text{Re} \, \mathcal{A}_{\ell m}, & m>0.
	\end{cases}
	\quad\quad 
	\mathcal{A}_{\ell m}=
\begin{cases}
		\frac{(-1)^m}{\sqrt{2}}\left(a_{\ell -m}+i a_{\ell m}\right), & m<0 \\
		a_{\ell m }, & m=0 \\
		\frac{1}{\sqrt{2}}\left(a_{\ell m}-i a_{\ell -m}\right), & m>0,
	\end{cases}
\end{equation}
where we have assumed that the function on the sphere being expanded is real.  The transformation (\ref{eq:sh}) is unitary, that is, we can write
\begin{equation}
	Y_{\ell m}=\sum_{\bar{m}} U_{m, \bar{m}} \, \mathcal{Y}_{\ell\bar{m}},
\end{equation}
with $U$ a unitary matrix, whose matrix elements are implicitly defined by equation (\ref{eq:sh}). Because of the unitary transformation, real spherical harmonics are  orthonormal,
\begin{equation}
	\int d^2 \hat{n}\,  Y_{\ell_1 m_1}Y_{\ell_2 m_2}=\delta_{\ell_1\ell_2}\delta_{m_1 m_2},
\end{equation}
and they also  satisfy the addition theorem
\begin{equation}\label{eq:addition theorem}
	P_\ell(\hat{n}_1 \cdot \hat{n}_2)=\frac{4\pi}{2\ell+1}\sum_m Y_{\ell m}(\hat{n}_1) Y_{\ell m}(\hat{n}_2),
\end{equation}
where $P_\ell$ is a Legendre polynomial.

Sometimes we need to integrate over the product of three spherical harmonics. We define
\begin{equation}\label{eq:D}
	D(\ell_1, m_1; \ell_2, m_2; \ell_3, m_3)\equiv \sqrt{4\pi}\int d^2\hat{n} \, Y_{\ell_1 m_1} Y_{\ell_2 m_2} Y_{\ell_3 m_3},
\end{equation}
which clearly is totally symmetric in its three arguments. Since the real spherical harmonics are related to  the complex spherical harmonics by a unitary transformation, this expression is closely related to the integral of the product of three complex spherical harmonics $\mathcal{D}$. The latter can be expressed as a product of Clebsch-Gordan coefficients (or  Wigner symbols), so we have
\begin{equation}
	D(\ell_1, m_1; \ell_2, m_2; \ell_3, m_3)=\sum_{m_1, m_2, m_3}
	U_{m_1 \bar{m}_1} U_{m_2 \bar{m}_2}
	\mathcal{D}(\ell_1, \bar{m}_1; \ell_2, \bar{m}_2 | \ell_3, \bar{m}_3) U^\dag_{\bar{m}_3 m_3}, 
\end{equation}
with
\begin{equation}
\mathcal{D}(\ell_1, m_1; \ell_2, m_2 | \ell_3, m_3)=
	\sqrt{\frac{(2\ell_1+1)(2\ell_2+1)}{2\ell_3+1}}
	\langle \ell_1, 0; \ell_2, 0 | \ell_3, 0\rangle
	\langle \ell_1, m_1; \ell_2, m_2 | \ell_3, m_3\rangle. 
\end{equation}
It follows then for instance that  $\mathcal{D}(\ell_1, m_1; \ell_2, m_2 | \ell_3, m_3)=\mathcal{D}(\ell_1, -m_1; \ell_2, -m_2 | \ell_3, -m_3)$.

Under  (active) azimuthal rotations by an angle $\phi$ the complex spherical harmonic coefficients transform according to $\mathcal{A}_{\ell m}\to e^{-i m\phi} \mathcal{A}_{\ell m}$. Therefore, it follows from the left equation in (\ref{eq:relations}) that real spherical harmonic coefficients $a_{\ell m}$ transform according to 
\begin{equation}\label{eq:M rotation}
	a_{\ell m}\to \cos(m\,\phi) a_{\ell m}-\sin(m\,\phi) a_{\ell -m}.
\end{equation}

\end{document}